\RequirePackage{fix-cm}
\documentclass[smallextended]{svjour3}       
\smartqed  
\usepackage{graphicx}
\usepackage{amsmath}
\usepackage{breqn}
\usepackage{multirow}
\begin{document}

\title{A principled distance-based prior for the shape of the Weibull model
}
%



\author{Janet van Niekerk\and Haakon Bakka\and H{\aa}vard Rue} 


\institute{J. van Niekerk \at
              CEMSE Division, King Abdullah University of Science and Technology, Saudi Arabia \\
              Tel.: +966-12-8088215\\
              \email{Janet.vanNiekerk@kaust.edu.sa}           
           \and
           H. Bakka \at
              CEMSE Division, King Abdullah University of Science and Technology, Saudi Arabia 
              \and 
              H. Rue \at
              CEMSE Division, King Abdullah University of Science and Technology, Saudi Arabia 
}
\date{Received: date / Accepted: date}

\maketitle

\begin{abstract}
The use of flat or weakly informative priors is popular due to the objective a priori belief in the absence of strong prior information. In the case of the Weibull model the improper uniform, equal parameter gamma and joint Jeffrey's priors for the shape parameter are popular choices. The effects and behaviours of these priors have yet to be established from a modeling viewpoint, especially their ability to reduce to the simpler exponential model. In this work we propose a new principled prior for the shape parameter of the Weibull model, originating from a prior on the distance function, and advocate this new prior as a principled choice in the absence of strong prior information. This new prior can then be used in models with a Weibull modeling component, like competing risks, joint and spatial models, to mention a few. This prior is available in the \textit{R-INLA} for use, and is applied in a joint longitudinal-survival model framework using the INLA method.
\end{abstract}

\section{Introduction}
\label{intro}
The Weibull model \cite{weibull1951} is a generalization of the exponential model \cite{kondo1930,weibull1939} through the inclusion of a positive shape parameter. In reliability or survival analysis, this venture allows the hazard function to change over time, as opposed to the constant hazard function from the exponential model. The Weibull model is still popular in many applied sciences, including reliability and survival analysis \cite{ataee2018,erango2019,brandl2020influence,chilikov2020,marin2005}, wind speed modeling \cite{shu2015,de2019,galarza2019assessment} and quality engineering \cite{tsiamyrtzis2006,colosimo2008}, to mention a few. However, this flexibility should also propel the cautious estimation of the shape parameter. An estimation method should be based on clear and transparent principles that are easily communicated \cite{gelman2017}. This ensures that the analyst conducts the modeling in a principled way and can incorporate useful prior information. Especially in reliability analysis or survival data, objectivity is improbable since the analyst will have some inherent subjectivity based on the subjects' nature and their generated data, being it a mechanical system or the physiological process of some disease. This prior information (or lack thereof) should be used to establish a principled estimation method for the shape parameter, which is what we propose here. \\ \\
The shape of the hazard function in a survival or reliability study describes the instantaneous risk of the event over time. If the hazard is constant over time, i.e. the exponential model is appropriate, then the hazard is equal at any point in time. However, if the hazard function is not constant then certain practical consequences will follow as a result. \\
For example, in a study on aggressive cancer, the time until relapse might be of interest. If the hazard function is constant then the follow-up times could be equally spaced as to minimize the time and cost commitments. If the hazard function is not constant, then follow-up times should either be condensed (increasing hazard) or prolonged (decreasing hazard), to be economically efficient.\\ The shape of the hazard function thus plays a pivotal role in the use of the model. We would thus aim to estimate the shape based on principles, and with a good understanding of the estimation process. In the case of constant hazard, we want to be able to recover the exponential model from the Weibull model. However, most current priors do not encourage the shrinkage to the exponential model or provide a mechanism to quantify the a priori belief about the strength of this shrinkage. \\ \\
Since the work of \cite{soland1969}, various priors for the shape parameter has been proposed.  Traditionally, assigning a prior to the shape parameter has been done through defining or deriving a distribution for the shape parameter itself. Often, an interpretable function of a hyperparameter has been proven to be a more natural parameter to estimate, and incorporate prior information for. A $\log$ parameterization of the shape parameter could be useful with a mode at zero (which implies the exponential model), although the contraction towards zero would be hard to quantify. Another issue is the (a)symmetry of the prior around the exponential model. Do we want (a)symmetric priors for the shape parameter, why and in what sense should it be asymmetric? \\ \\
Improper or noninformative priors have been popular for some time, but the need for weakly informative priors has been realised \cite{lemoine2019}. The improper uniform and the gamma priors have been used in various studies for the shape of the Weibull model, see \cite{abrams1996}, \cite{erango2019}, \cite{ataee2018}, \cite{marin2005}, \cite{chaturvedi2014} and \cite{gupta2017} amongst others. A joint Jeffrey's prior proposed by \cite{guure2012,gupta2017} is also a popular choice amongst practitioners \cite{aslam2014,gupta2017}. We will show in this paper that the use of these priors, without strong prior information, could lead to unreliable results and thus misspecified models. This type of result is not surprising and has been presented by \cite{simpson2017} and \cite{klein2016}, for different model setups. Our aim in this paper, is to propose a prior that provides reliable and predictable results in the Bayesian inference of the Weibull model and can quantify the contraction to the exponential model. Invariance under reparameterizations of the shape parameter would be beneficial, as is not the case for most popular priors.\\ \\
We propose a prior that is based on a distance metric in a principled and transparent way, from an information theoretic perspective. We use the work of \cite{simpson2017} to derive the penalized complexity (PC) prior for the shape parameter of the Weibull distribution. Even though our results are valid for all Weibull models, we provide the details for survival data. In Section \ref{secweibull} the Bayesian Weibull regression (survival) model is introduced and some of its properties are discussed. Then in Section \ref{secpc}, we derive and define the PC prior for the shape parameter and in Section \ref{secpcmot} discuss the effect of some popular priors on the model estimation, and posit the need for the consideration of the PC prior as a reliable prior. We conclude the paper with an application of the PC prior in a joint survival-longitudinal model setting in Section \ref{secapp}, and a discussion in Section \ref{secdisc}.
\section{Bayesian Weibull regression model for survival data}\label{secweibull}
\begin{figure}[h]
    \centering
    \includegraphics[width=5cm]{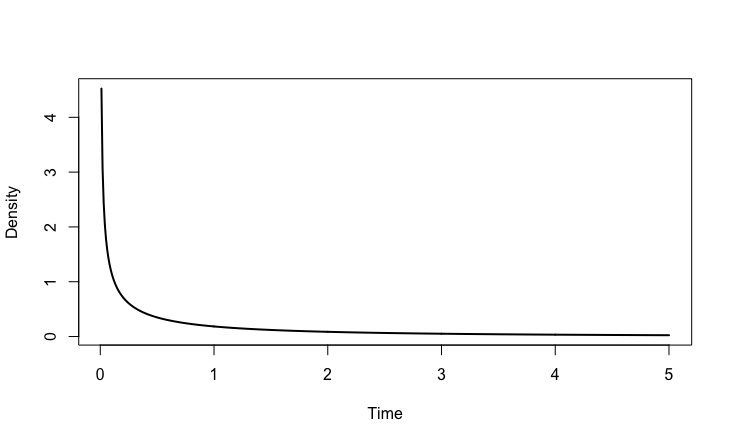}
     \includegraphics[width=5cm]{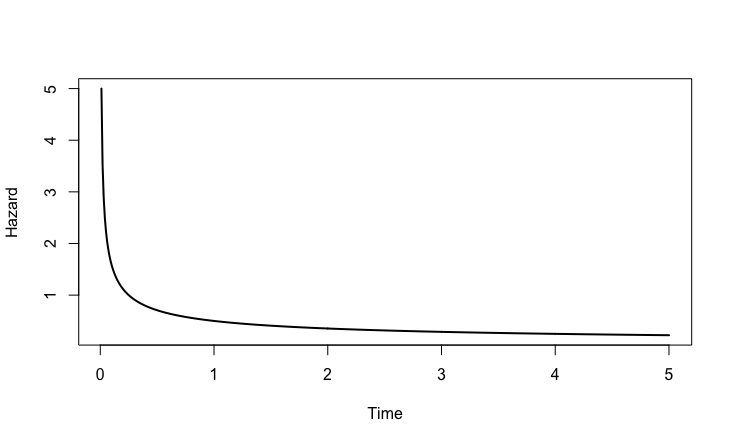}\\
     \includegraphics[width=5cm]{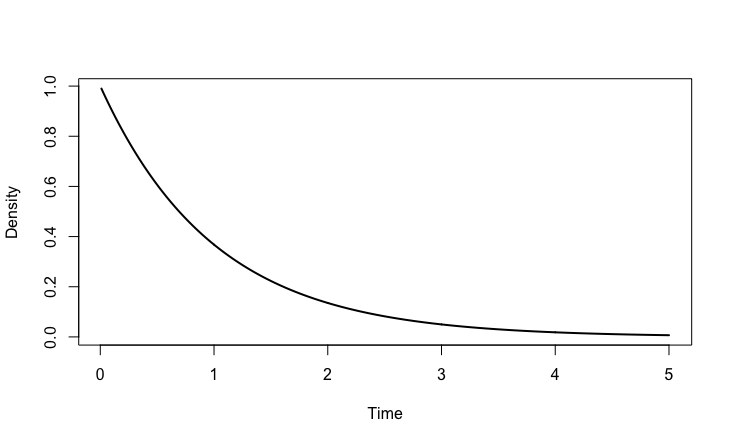}
      \includegraphics[width=5cm]{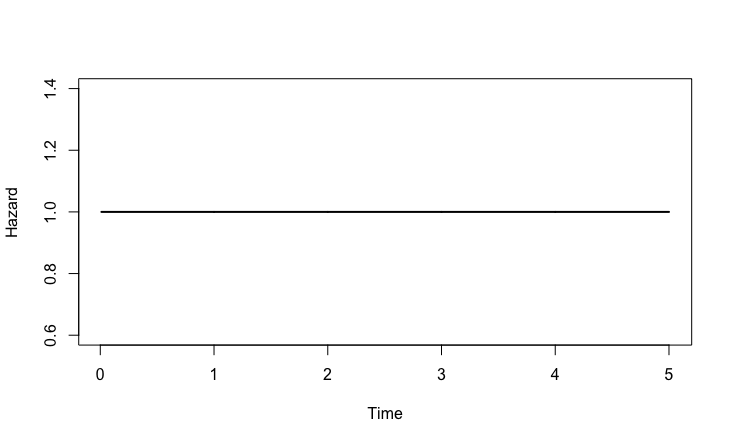}\\
      \includegraphics[width=5cm]{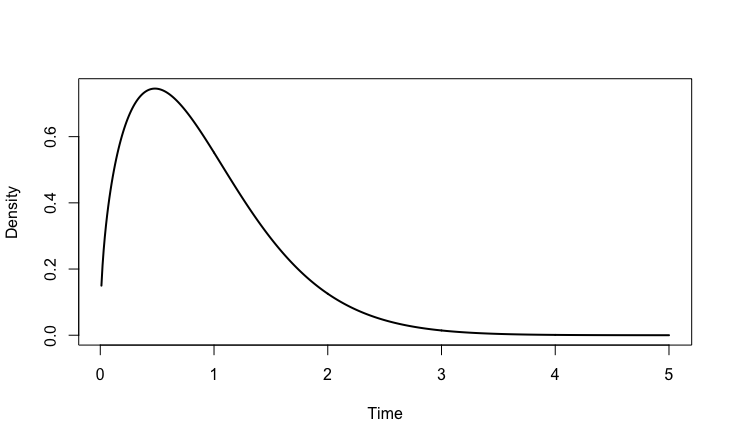}
        \includegraphics[width=5cm]{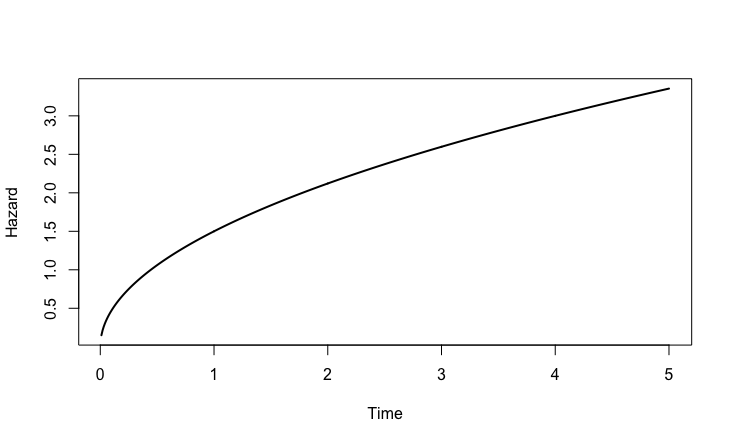}
    \caption{Densities (left) and hazard functions (right) of the Weibull model \eqref{par1} and \eqref{par2} for $\alpha=0.5,1,1.5$ (from top to bottom) respectively.}
    \label{figalpha}
\end{figure}

We focus our attention on the Weibull model in the context of survival data of the remainder of this paper. There are two parameterizations for the density function of a Weibull distributed random variable $Y$, with shape and scale parameters $\alpha$ and $\lambda$, respectively, as follows,
\begin{equation}
    f_1(y|\alpha,\lambda)=\alpha y^{\alpha-1}\lambda\exp(-\lambda y^\alpha), \quad y\geq 0,\alpha>0,\lambda>0
    \label{par1}
\end{equation}
and
\begin{equation}
    f_2(y|\alpha,\lambda)=\frac{\alpha}{\lambda}\left(\frac{y}{\lambda}\right)^{\alpha-1}\exp\left(-\left(\frac{y}{\lambda}\right)^\alpha\right),\quad y\geq 0,\alpha>0,\lambda>0.
    \label{par2}
\end{equation}
The parameterization in \eqref{par1} is still popular in practice but we will show that for orthogonal interpretation of the parameters, the parameterization in \eqref{par2} is essential.\\ \\
In survival analysis, the hazard function, $h(y)$ is a useful measure to understand the instantaneous risk of the event. For the Weibull model in \eqref{par1} the hazard function is
\begin{equation*}
    h(y)=\frac{f(y)}{S(y)}=\alpha\lambda y^{\alpha-1},
\end{equation*}
where $S(y)=\int_y^\infty f(u)du$, and for \eqref{par2}
\begin{equation*}
    h(y)=\alpha\lambda^{-\alpha} y^{\alpha-1}.
\end{equation*}
The effect of the value of $\alpha$ is illustrated in Figure \ref{figalpha} for a constant scale $\lambda=1$, without loss of generality. In this case the two parameterizations \eqref{par1} and \eqref{par2} lead to identical hazard functions. The flexibility of the Weibull model imposed by $\alpha$ is clear from Figure \ref{figalpha}. We note that decreasing, constant (exponential model) and increasing hazard functions are possible with this model. The type of hazard is determined by the value of $\alpha$ and care should thus be taken in the estimation of $\alpha$.
\subsection{Estimation of parameters}
In the presence of covariates, covariate information is incorporated in $\pmb{X}$ and then we define the scale parameter in \eqref{par1} or \eqref{par2} as $\lambda=\exp(\pmb{\beta X})$. A prior distribution is then explicitly formulated for $\pmb\beta$ instead of $\lambda$. Suppose the joint prior for $\pmb\beta$ and $\alpha$ is represented by $\pi(\pmb\beta,\alpha)$. \\
The likelihood function for a sample of $N$ individuals with non-censoring indicators $d_i,i=1,...,N$ (i.e. $d_i=1$ if $y_i$ is a complete observation) and covariates $\pmb{X}$, is
\begin{equation}
    L(\alpha,\pmb\beta|\pmb y, \pmb d)=\prod_{i=1}^{N}f(y_i|\alpha,\pmb\beta)^{d_i}S(y_i|\alpha,\pmb\beta)^{(1-d_i)},
    \label{likbeta}
\end{equation}
where $f(y_i|\alpha,\pmb\beta)$ is from \eqref{par1} or \eqref{par2} and $S(y_i|\alpha,\pmb\beta)=\int_{y_i}^\infty f(u|\alpha,\pmb\beta)du$.
Censored observations lead to the inclusion of the survival function $S(y_i|\alpha,\pmb\beta)$ in the likelihood function to account for incomplete observations, which is unique to the survival Weibull model, as opposed to a Weibull model for other strictly positive responses where all observations are observed completely. \\ \\
The prior and the likelihood information is then combined to form the joint posterior, $\pi(\alpha,\pmb\beta|\pmb{y},\pmb{d})\propto L(\alpha,\pmb\beta|\pmb y, \pmb d)\pi(\alpha, \pmb\beta)$, after which Bayesian inference is possible. Often, also here, the explicit form of the posterior density is not analytically tractable and scientists revert to simulation-based methods for Bayesian inference, like Markov chain Monte Carlo (MCMC) methods \cite{brooks2011} or approximate methods like integrated nested Laplace approximations (INLA) \cite{rue2009}.

\subsection{Popular prior choices}\label{seccurrentpriors}
\begin{figure}[h]
\includegraphics[width=12cm]{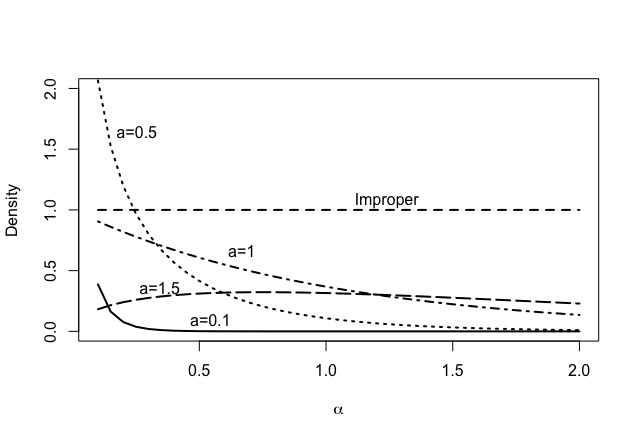}
\caption{Improper and Gamma$(a,a)$ prior densities for $\alpha$ on the $\alpha$ scale}
\label{figprioralp}
\end{figure}

There is no joint conjugate prior for $(\alpha,\lambda)$ (or $(\alpha,\pmb\beta)$) if all parameters are assumed to be unknown. It is common to assign independent priors to $\alpha$ and $\pmb\beta$ \cite{banerjee2008,chaturvedi2014,ataee2018,gupta2017}, respectively or use the joint Jeffrey's prior \cite{guure2012,gupta2017}.\\ \\
In our current work, we consider the Gaussian prior for $\pmb{\beta}$ and mainly focus on the prior for the shape parameter, $\alpha$.
The two most common priors of $\alpha$ are thus:
\begin{enumerate}
    \item {\textbf{Improper uniform prior.}\\
    The improper uniform prior translates into a constant density for all values of $\alpha$, i.e. \begin{equation}
        \pi_I(\alpha)\propto 1.
        \label{improperpriordensity}
    \end{equation}}
    \item {\textbf{Gamma prior.}\\
    A gamma prior with equal parameters is assigned to $\alpha$ i.e. $\alpha \sim Gamma(a,a)$ such that the prior density function of $\alpha$ is given by
    \begin{equation}
        \pi_G(\alpha)=\frac{a^{a}}{\Gamma(\alpha)}\alpha^{a-1}\exp(-a \alpha).
        \label{gammapriordensity}
    \end{equation}}
\end{enumerate}
The resulting priors \eqref{improperpriordensity} and \eqref{gammapriordensity} for some $a$ values, are illustrated in Figure \ref{figprioralp}. Even though all the gamma priors have a priori $E(\alpha)=1$, it is clear from Figure \ref{figprioralp} that the mode of the priors is not at $\alpha=1$, or in the vicinity. Actually, the density at $\alpha=1$ is small for most $a$ values, which results in a prior belief that the base model is very unlikely, even though the prior expected value indicates the base model. As a default prior choice, this is undesirable. Only if expert knowledge justifies the gamma prior to exhibit some particular behaviour of the parameter, the gamma prior might be appropriate. \\ \\ 
This leaves the user with the challenge to formulate a default prior for $\alpha$ which will contract to the base model value at a quantifiable rate, since neither the improper or gamma priors satisfies this desideratum. It is thus our aim in this paper to propose a principled prior based on the distance to the base model $(\alpha=1)$ rather than just assigning certain prior density values to arbitrary $\alpha$ values. This is achieved in Section \ref{secpc}, particularly in Section \ref{secpcweibull}.

\section{Penalized complexity prior}\label{secpc}
Penalized complexity priors (PC priors) as defined by \cite{simpson2017} have been shown to be principled and sensible prior choices for hyperparameters where the definition of a base model is natural. Currently, the \textit{R-INLA} software available in the \textit{INLA} library in \textit{R} is the only computational tool that contains the PC priors as prior options for various modeling elements. In this section we firstly present an overview of the methodology that underpins the PC priors and then we develop the PC prior for the shape of the Weibull model.
\subsection{Methodology}
The PC prior is derived from the Kullback-Leibler divergence (KLD,\cite{kullback1951}) between the proposed (complex) model and a natural base model. In the context of the Weibull model, the natural base model is the exponential model, where $\alpha=1$ in \eqref{par1} and \eqref{par2}. \\ \\
The KLD between two density functions $f$ and $g$ for a random variable $y$, is defined as 
\begin{equation}
    \text{KLD}(f(y)||g(y))=\int_Df(y)\log\left(\frac{f(y)}{g(y)}\right)dy,
\end{equation}
where $D$ is the support of $y$. The KLD is a measure of the information lost when choosing $g(y)$ over $f(y)$. Now in our case, let $f_i(y)$ be the complex model and $g(y)$ be the base (exponential) model, i.e. $g(y)=f_i(y|\alpha=1)$, then the KLD for either parameterization $(i=1,2)$ is defined as
\begin{equation}
   \text{KLD}_i(f_i||g)=\text{KLD}_i(\alpha)=\int_0^\infty f_i(y)\log\left(\frac{f_i(y)}{g(y)}\right)dy,
\end{equation}
for $i=1,2$ from \eqref{par1} or \eqref{par2}. Then define the distance function $$d_i(\alpha)=\sqrt{2\text{KLD}_i(\alpha)}.$$ This distance function is a measure of the complexity of the more complex model with respect to the exponential model, in the sense that it is a unidirectional distance between the complex model $f_i(y)$ and the exponential model $g(y)$. The PC prior is then constructed through $d_i(\alpha)\sim Exp(\theta)$ and hence,
\begin{equation*}
    \pi(\alpha)=\theta\exp\left[-\theta d_i(\alpha)\right]\left|\frac{\partial d_i(\alpha)}{\partial \alpha}\right|,
\end{equation*}
where $\theta$ is a user-defined hyperparameter such that $P(d_i(\alpha)>U)=p,p>0$ where $p$ is small. This statement incorporates the information of the tail behaviour of the prior, based on the information from the user.
\subsection{PC prior for both parameterizations}
In this section we derive the KLD and subsequently the PC prior for $\alpha$ for both parameterizations \eqref{par1} and \eqref{par2}. Suppose that the information from covariates are incorporated in $\lambda$ for the complex model $(\alpha \neq 1)$ and in $\lambda_0$ for the base model $(\alpha=1)$ as described in Section \ref{secweibull}.
\subsubsection{Parameterization 1}
From \eqref{par1} the KLD for the first parameterization is
\begin{eqnarray*}
&&\text{KLD}_1(\alpha|\lambda,\lambda_0)\\
&=&\text{KLD}\left(f(y|\alpha,\lambda)||f(y|\alpha=1,\lambda_0)\right)\notag\\
&=&\int_{0}^{\infty}f(y|\alpha,\lambda)\log\left(\frac{f(y|\alpha,\lambda)}{f(y|\alpha=1,\lambda_0)}\right)dy\notag\\
&=&\alpha^{-1}\left(\gamma-\alpha(1+\gamma)+\lambda^{-\frac{1}{\alpha}}\lambda_0\Gamma\left(\frac{1}{\alpha}\right)+\alpha\log(\alpha)+\log(\lambda)-\alpha\log(\lambda_0)\right),\quad\label{kldpar1}
\end{eqnarray*}
where $\gamma=0.577216$ is Euler's constant (see \cite{euler1735}). Furthermore, suppose $\lambda_0=\lambda$ (which is a realistic assumption since the covariates will be the same for an individual no matter the model) then 
\begin{equation}
\text{KLD}_1(\alpha|\lambda)=\alpha^{-1}\left(\gamma-\alpha(1+\gamma)+\lambda^{1-\frac{1}{\alpha}}\Gamma\left(\frac{1}{\alpha}\right)+\alpha\log(\alpha)+(1-\alpha)\log(\lambda)\right).
\label{kldpar1}
\end{equation}
It is clear from \eqref{kldpar1} that the KLD for $\alpha$ is dependent on the value of $\lambda$. This is undesirable and impractical and we will thus not use this parameterization.


\subsubsection{Parameterization 2}
The KLD for the second parameterization from \eqref{par2} is
\begin{eqnarray*}
&&\text{KLD}_2(\alpha|\lambda,\lambda_0)\\
&=&\text{KLD}\left(f(y|\alpha,\lambda)||f(y|\alpha=1,\lambda_0)\right)\notag\\
&=&\int_{0}^{\infty}f(y|\alpha,\lambda)\log\left(\frac{f(y|\alpha,\lambda)}{f(y|\alpha=1,\lambda_0)}\right)dy\notag\\
&=&\alpha^{-1}\left(\gamma-\alpha(1+\gamma)+\frac{\lambda}{\lambda_0}\Gamma\left(\frac{1}{\alpha}\right)+\alpha\log(\alpha)+\alpha\log\left(\frac{1}{\lambda}\right)-\alpha\log\left(\frac{1}{\lambda_0}\right)\right),\notag
\\&&\label{kldpar2}
\end{eqnarray*}
where $\gamma=0.577216$ is Euler's constant (see \cite{euler1735}). Now, again suppose that $\lambda=\lambda_0$, then 
\begin{equation}
\text{KLD}_2(\alpha|\lambda)=\alpha^{-1}\left(\gamma-\alpha(1+\gamma)+\Gamma\left(\frac{1}{\alpha}\right)+\alpha\log(\alpha)\right).
\label{kldpar2}
\end{equation}
Now, from \eqref{kldpar2} we can see that the PC prior for $\alpha$ is independent of the value for $\lambda$. This result advocates the use of the second parameterization as the Weibull model. \\ \\
Subsequently the PC prior is derived from the KLD as
\begin{eqnarray}
\pi(\alpha)=\theta\exp\left[-\theta d(\alpha)\right]\left|\frac{\partial d(\alpha)}{\partial \alpha}\right|,
\label{pc2}
\end{eqnarray}
where $\theta$ is such that $P(d(\alpha)>U)=p$, i.e. $\theta=-\frac{\ln p}{U}$. The choice of $\theta$ gives the user control over the tail behaviour of the prior and hence the rate of contraction towards the base model. We show the influence of $\theta$ in the next section.
\subsection{PC prior for the shape parameter of the Weibull model}\label{secpcweibull}
\begin{figure}[h]
    \centering
    \includegraphics[width=12cm]{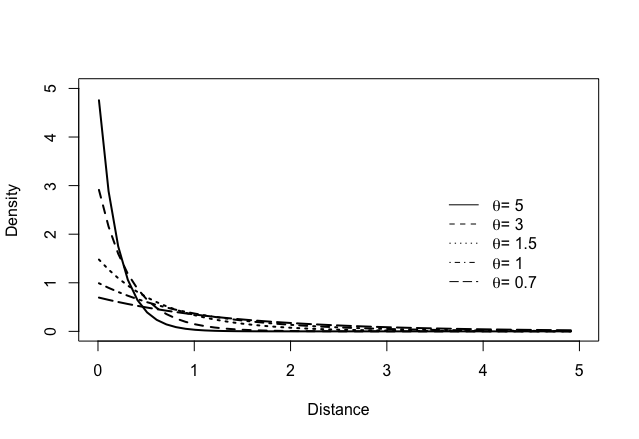}
  \caption{Penalizing complexity prior density for $\alpha$ as in \eqref{pcweibull} on the distance to the base model $(\alpha=1)$ scale}
    \label{figpc_dist}
\end{figure}

\begin{figure}[h]
    \centering
    \includegraphics[width=12cm]{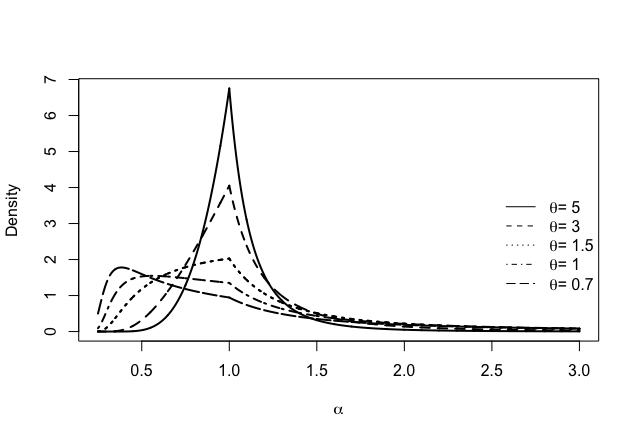}
  \caption{Penalizing complexity prior density for $\alpha$ as in \eqref{pcweibull} on the $\alpha$ scale}
    \label{figpc_alpha}
\end{figure}

As discussed in the previous subsection we will only consider parameterization 2 of the Weibull model as in \eqref{par2} for the remainder of this paper. From \eqref{kldpar2} and \eqref{pc2} the PC prior is defined as 
\begin{eqnarray}
\pi(\alpha)&=&\theta\exp\left[-\theta\sqrt{2\text{KLD}_2(\alpha)}\right]\left|\frac{\partial\sqrt{2\text{KLD}_2(\alpha)}}{\partial \alpha}\right|\notag \\
&=&\theta\exp\left[-\theta\sqrt{2\text{KLD}_2(\alpha)}\right]\notag \\
&&\times \frac{1}{2}\left(2\text{KLD}_2(\alpha)\right)^{-\frac{1}{2}}\left|\frac{2}{\alpha^2}\left(\gamma-\alpha(1+\gamma)+\Gamma(\alpha^{-1})+\alpha\log\alpha\right)\vphantom{...}\right.\notag\\
&&\left.\vphantom{...}+\frac{2}{\alpha}\left(-(1+\gamma)-\frac{\Gamma(\alpha^{-1})\psi(\alpha^{-1})}{\alpha^2}+\log\alpha+1\right)\right|
\label{pcweibull}
\end{eqnarray}
where $\psi(z)$ is the digamma function \cite{abramowitz1988}.\\ \\
In Figure \ref{figpc_dist} we present the PC priors on the distance scale for different values of $\theta$. Similarly we present the PC priors on the $\alpha$ scale in Figure \ref{figpc_alpha}.
\\ \\
The exponential prior for the distance to the base model is clear from Figure \ref{figpc_dist}. We observe that smaller values of $\theta$ have weaker contraction to the base model as a result of the heavier exponential tails. 
The PC priors are shown in Figure \ref{figpc_alpha} on the original $\alpha$ scale. This figure has some interesting features. It is clear that the modal density is at the base model $(\alpha=1)$ for all $\theta> 1$, but not for $\theta\leq 1$. This is contradictory to what we expect in the PC prior in the sense that the modal density on the original $\alpha$ scale might not be at $\alpha=1$, although the PC prior still has its mode at a distance of zero. Some of the resulting shapes of the PC prior $(\theta<1)$ on the $\alpha$ scale is quite unexpected and shows how important it is to define a prior on a meaningful invariant scale, based on principles and transparency. Surprisingly, a mode at $\alpha=1$ is not necessary for a good principled prior.  \\ \\
Also, as showed by \cite{simpson2017}, the PC prior is invariant under transformations of the shape parameter. This invariance is necessary for proper inference with consistent results amongst reparameterizations of $\alpha$.\\ \\
In the next section we investigate some popular priors and illustrate their differences to the PC prior \eqref{pcweibull}.

\section{The PC prior as a principled prior for $\alpha$}\label{secpcmot}
In this section we advocate the use of the proposed PC prior as a default prior when the user does not have strong enough prior information to formulate another prior. In Sections \ref{intro} and \ref{seccurrentpriors} we hinted to the challenges presented by some of the popular prior options and now we present these challenges from the perspective of the distance to the base model, i.e. the exponential model.

\subsection{Performance of the priors on the distance scale}
To properly understand the behaviour of the other priors we need to reformulate these priors; from a prior in terms of $\alpha$ to a prior in terms of the distance between the proposed Weibull model that stems from the chosen prior, to the exponential base model where $\alpha=1$. To this end we will again use the distance measure $\sqrt{2\text{KLD}_2(\alpha)}$ from \eqref{kldpar2}. We present the gamma prior here for brevity but the same can be done for the other priors. \\
We need to reparameterize the gamma prior \eqref{gammapriordensity}, $\pi_G(\alpha)$, in terms of the distance, to find the corresponding prior densities based on the distance scale. Analytically, this endeavour is time-consuming and of little value so we investigate this numerically. Some distance values to the base model, their corresponding $\alpha$ values and densities from the gamma prior \eqref{gammapriordensity} are presented in Tables \ref{tablepriorsdist} and \ref{tablepriorsdist_0.1}. \\

\begin{figure}[h]
    \centering
    \includegraphics[width=5.5cm]{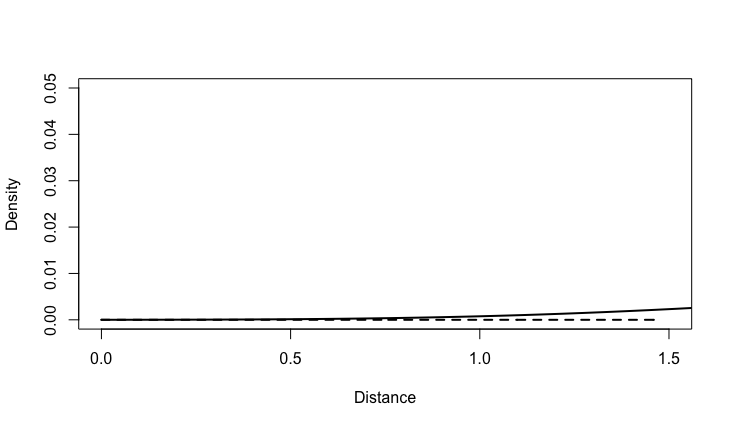}
    \includegraphics[width=5.5cm]{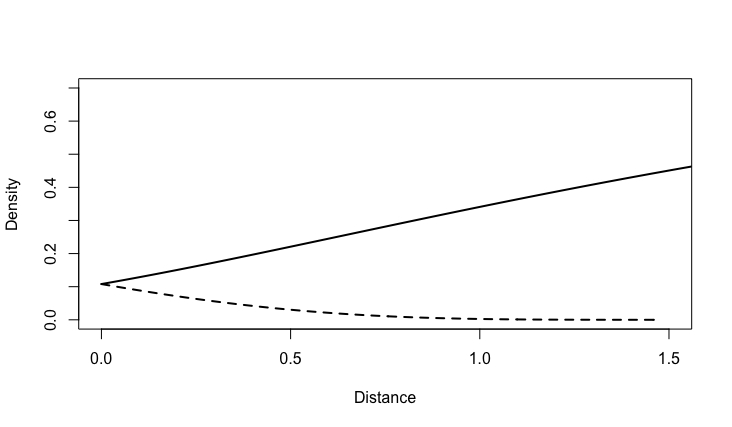}\\
    \includegraphics[width=5.5cm]{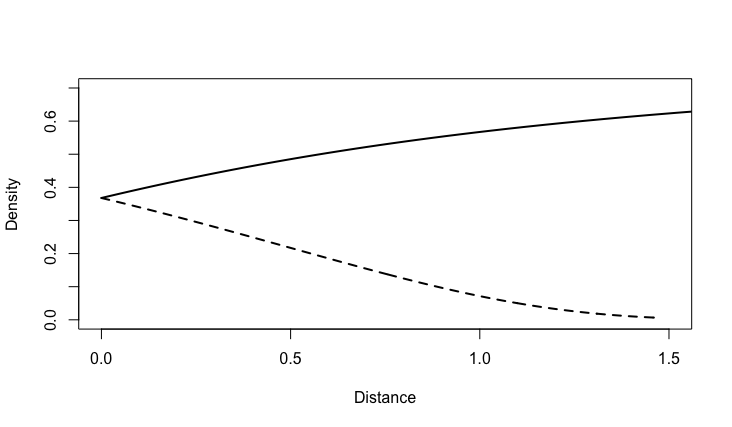}
    \includegraphics[width=5.5cm]{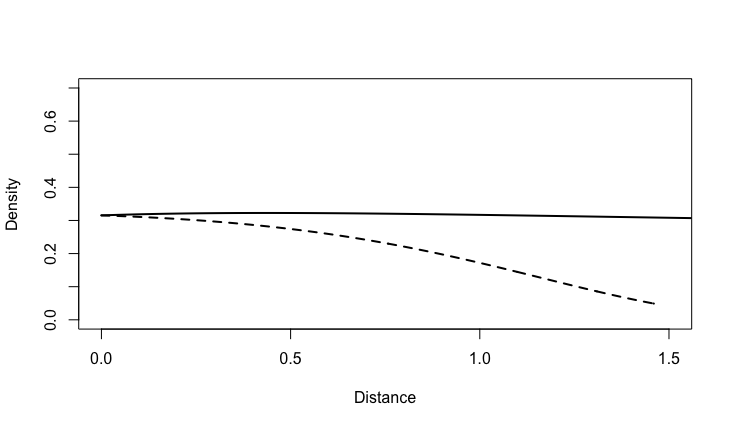}
    \caption{Gamma$(a,a)$ prior densities for $\alpha$ on the distance scale ($a=0.1,a=0.5,a=1,a=1.5$ from top left to bottom right with a solid curve for $\alpha\leq1$ and dashed for $\alpha\geq 1$).}
    \label{figpriordist}
\end{figure}

\begin{table}[h]
\caption{Some distance values and corresponding $\alpha$ and $\pi_G(\alpha)$ values for $a=1.5$.}
\label{tablepriorsdist}       
\begin{tabular}{lll}
\hline\noalign{\smallskip}
Distance & $\alpha\leq1$ value and $\pi_G(\alpha)$ & $\alpha\geq1$ value and $\pi_G(\alpha)$\\
\noalign{\smallskip}\hline\noalign{\smallskip}
$0$ & $1.00\quad (0.315)$ & $1.00\quad (0.315)$ \\
$0.1$ & $0.93\quad (0.319)$ & $1.08\quad (0.311)$ \\
$0.5$ & $0.72\quad (0.322) $ & $1.53\quad (0.274)$\\
$0.8$ & $0.62\quad (0.320) $ & $2.09\quad (0.220)$\\
$1.45$ & $0.48\quad (0.309) $ & $4.93\quad (0.051)$\\
\noalign{\smallskip}\hline
\end{tabular}
\end{table}
\noindent From Table \ref{tablepriorsdist} we can see that for each positive distance value, there are two different density values, each one corresponding to a value of $\alpha\leq1$ and $\alpha\geq1$, respectively. These two values of $\alpha$ are not equidistant. This implies that the Gamma$(1.5,1.5)$ prior results in different contractions to the base model for equidistant values of $\alpha$ and $1$. This discriminating behaviour is again evident in Table \ref{tablepriorsdist_0.1} for the Gamma$(0.1,0.1)$ prior. This prior is still commonly used in the literature \cite{chaturvedi2014,gupta2017} maybe because it seems to be quite uninformative from Figure \ref{figprioralp}, but in Table \ref{tablepriorsdist_0.1} we can see clearly that this is not a good option without proper justification since the density at the base model (distance of zero) is very close to zero, and the rate of the discrimination will be very hard to motivate.
\begin{table}[h]
\caption{Some distance values and corresponding $\alpha$ and $\pi_G(\alpha)$ values for $a=0.1$.}
\label{tablepriorsdist_0.1}       
\begin{tabular}{lll}
\hline\noalign{\smallskip}
Distance & $\alpha\leq1$ value and $\pi_G(\alpha)$ & $\alpha\geq1$ value and $\pi_G(\alpha)$\\
\noalign{\smallskip}\hline\noalign{\smallskip}
$0$ & $1.00\quad (<0.0001)$ & $1.00\quad (<0.0001)$ \\
$0.1$ & $0.93\quad (<0.0001)$ & $1.08\quad (<0.0001)$ \\
$0.5$ & $0.72\quad (<0.0001) $ & $1.53\quad (<0.0001)$\\
$0.8$ & $0.62\quad (0.0004) $ & $2.09\quad (<<0.0001)$\\
$1.45$ & $0.48\quad (0.002) $ & $4.93\quad (<<0.0001)$\\
\noalign{\smallskip}\hline
\end{tabular}
\end{table}
We present these results graphically in Figure \ref{figpriordist}. The modal a priori density is not at the base model but some other "random" configuration. The discriminating behaviour of the Gamma prior for values smaller or larger than $1$ as shown in Tables \ref{tablepriorsdist} and \ref{tablepriorsdist_0.1}, is very clear from Figure \ref{figpriordist}. We envision that it would be very hard to properly motivate such a discrimination in terms of the distance to the base model and this shows clearly why the Gamma$(a,a)$ prior should not be used without prior justification, even for $a=0.1$. 
\subsection{Computational considerations for the PC prior}
The Integrated Nested Laplace Approximation (INLA) method was developed by \cite{rue2009} as a computational tool for the (approximate) Bayesian inference of latent Gaussian models. For more details on latent Gaussian models and INLA in the context of survival models, see \cite{rue2005} and \cite{martino2011} amongst others.\\ \\
The proposed PC prior \eqref{pcweibull} is currently implemented in the \textit{INLA} library in R through the \textit{pc.alphaw} option for the prior of the hyperparameter $\alpha$. Functions to evaluate, sample, compute quantiles and percentiles of this prior is also included in the current version of the \textit{INLA} library. \\
Currently, the default prior for the Weibull model is the PC prior with a parameter value $\theta=2.5$. Various other Bayesian software can be used for inference using the PC prior \eqref{pcweibull}.\\ \\

\section{Application}\label{secapp}
In this section we apply the proposed PC prior in the context of a survival model. In the preceding sections we motivated and proposed the PC prior by only considering classical survival models. This restriction is not necessary. The PC prior can be applied as a prior for the hyperparameter of the Weibull survival model in the context of competing risks models, joint survival-longitudinal models, multistate models, spatial survival models and many more. Here we show a joint survival-longitudinal application purely for illustration, as we could have simply presented various other models.\\ \\
In prostate cancer studies, Prostate-specific Antigen (PSA) has been
identified as a biomarker, measured at multiple timepoints, for the status of prostate cancer. High
levels of PSA are indicative of increased risk of prostate cancer or
recurrence. Radiation therapy is a common course of treatment often
prescribed for patients with prostate cancer. If successful, the PSA
levels are expected to drop and remain at a low level. If not, then PSA levels will still drop initially but then rise again
\cite{zagars1995}. The estimation of the longitudinal trajectory of PSA is further complicated by the dropout process whereby patients' are no longer part of the study, either due to salvage hormone therapy or recurrence of the cancer. This is known as informative drop-out, which in turn implies informative censoring of the data. If this informative drop-out is unaccounted for, it can lead to
considerable bias in the PSA trajectory estimation.\\ \\ 
The objective of
this application is thus to identify the trajectory of post-radiation
PSA change, while correctly accounting for the informative drop-out. To achieve this, a joint longitudinal-survival model is defined with the PSA levels as the longitudinal process and the time to informative drop-out as the survival process with the logarithm of the base PSA value as a linear covariate, $\log (\text{PSA}_{\text{base}})$ .
More details about the clinical impact of such a study can be found in
\cite{proust2009}.\\ \\
The joint model under consideration in this application is of the form:
\begin{eqnarray}
\log(\text{PSA})(t)&=&\eta^L(t)+\epsilon(t)\notag \\
h(t)&=&h_0(t)\exp(\eta^S(t))\notag
\end{eqnarray} 
where $\epsilon\sim N(0,\sigma^2_\epsilon)$, $\eta^L$ and $\eta^S$ are the longitudinal and dropout linear predictors, respectively. We assume a Weibull
hazard function for the dropout process. The exponential case with
constant hazard can be achieved as a special case when $\alpha=1$. The
linear predictors, $\eta^L$ and $\eta^S$, are formulated as:
\begin{eqnarray}
\eta^L(t)&=&\kappa(t)+\beta \log (\text{PSA}_{\text{base}})+w+vt\notag\\
\eta^S(t)&=&\gamma \log(\text{PSA}_{\text{base}})+\nu(w+vt)\label{surv1}
\end{eqnarray}
where
$$\begin{bmatrix} w \\ v \end{bmatrix}\sim
N \begin{pmatrix} \begin{bmatrix}0 \\ 0 \end{bmatrix}, \begin{bmatrix}
\sigma^2_w & \rho\sigma_w\sigma_v\\ \rho\sigma_w\sigma_v &
\sigma^2_v \end{bmatrix} \end{pmatrix}$$ and $\kappa(t)$ is a
Bayesian smoothing spline with hyperparameter $\sigma_{\kappa}$, that captures the non-linear temporal trend in the PSA level \cite{lindgren2008}.\\ \\
The priors for $\beta$ and $\gamma$ are Gaussian, and penalizing complexity priors as presented in \cite{simpson2017} are assigned to the precision/variance hyperparameters $\sigma_{\kappa}$, $\sigma_v$, $\sigma_w$, $\sigma_{\beta}$, $\sigma_{\gamma}$, $\sigma_{\nu}$, $\rho$ and the longitudinal model error standard deviation $\sigma_{\epsilon}$. Also, we now assume the PC prior for the shape parameter $\alpha$, proposed in this paper \eqref{pcweibull} (currently the default prior in \textit{R-INLA} for the Weibull model). We also include a sensitivity analysis using different values of $\theta$.\\ \\
\subsection{Results}\label{appresults}
\begin{figure}[h]
    \centering
    \includegraphics[width=12cm]{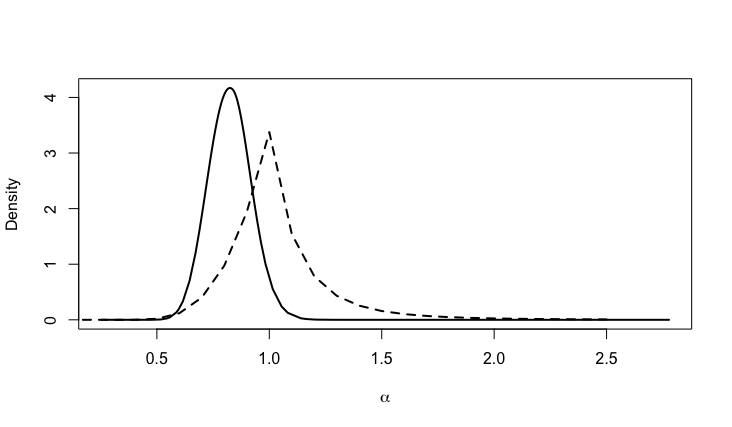}
    \caption{Posterior (solid) and prior (dashed) densities for $\alpha$.}
    \label{figPSA}
\end{figure}

The resulting estimated joint model is presented, with the PC prior with $\theta=2.5$ (the default value in \textit{R-INLA}). The
results are summarized in Table \ref{tableres1}.
\begin{table}[h]
	\centering
	\begin{tabular}{|c||c|c|}
		\hline
		\textbf{Parameter} & \textbf{Posterior Mode} & \textbf{Posterior SD}\\ \hline
		$\beta$ & $0.450$ & $0.061$\\
		$\gamma$ & $0.743$ & $0.198$ \\
		$\sigma^2_\epsilon$ & $0.091$ & $0.005$ \\
		$\sigma^2_\kappa$ & $0.226$ & $0.143$ \\
		$\sigma^2_w$ & $0.342$ & $0.053$ \\ 
		$\sigma^2_v$ & $0.216$ & $0.054$\\
		$\rho$ & $-0.131$ &$0.149$ \\ 
		$\nu$ & $0.921$ & $0.137$ \\
		$\alpha$ & $0.825$ & $0.094$ \\\hline  
	\end{tabular}
	\caption{Results for the PSA dataset}
	\label{tableres1}
\end{table}
It is quite clear from Table \ref{tableres1} that the hazard of
informative dropout is correlated with the longitudinal PSA biomarker
since $\nu=0.921$ with $95\%$ credible interval $(0.647;1.195)$. This
result confirms that the joint model approach is supported by the data
and should be preferred to the separate models. The structure of the
association term as in \eqref{surv1} is quite restrictive but has been
used extensively. We can assume various other association structures and do the inference using \textit{R-INLA}, but this is outside the scope of the current paper. \\ \\
The posterior and prior densities for $\alpha$ is presented in Figure \ref{figPSA}.  

\subsection{Sensitivity analysis}
\begin{figure}[h]
    \centering
    \includegraphics[width=12cm]{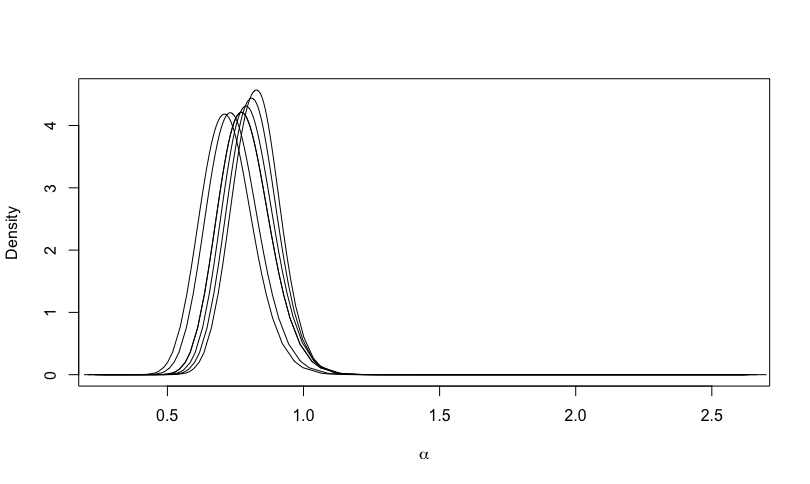}
    \caption{Marginal posterior densities of $\alpha$ for different values of $\theta$}
    \label{fig_ressens}
\end{figure}

In this section we redo the inference of Section \ref{appresults} for $\theta\in\{0.5,1,1.5,2,2.5,3,3.5,4,4.5,5\}$. Note that the value of $\theta$ influences the tail behaviour of the PC prior. As mentioned previously, a lower value of $\theta$ implies higher a priori probability for values away from $\alpha=1$, as in Figures \ref{figpc_dist} and \ref{figpc_alpha}.\\
We omit the tables similar to Table \ref{tableres1} since the results are very similar. In Figure \ref{fig_ressens} the marginal posterior density of $\alpha$ for the different $\theta$ values are presented, and the robustness of the PC prior is clear.

\section{Discussion}\label{secdisc}
Prior elicitation is one of the crucial elements in Bayesian analysis, and influences the inference substantially, especially if the data is not very informative. The proper elicitation of prior information, however, is often hard or not possible. In this case, popular or uninformative priors are reverted to since they have either been used extensively in the literature, or they are seemingly weakly/non informative. Here we did not debate informativeness or uninformativeness of priors, but rather the reliability, robustness and transparency of the prior. \\ In this paper we viewed the Weibull model as a generalization of the exponential model through the shape parameter, and as such the Weibull model is the complex counterpart of the exponential base model. This naturally leads to the provocation of a prior based on the distance from the more complex Weibull model to the base model, and thus the contraction to the base model. The penalizing complexity prior framework as proposed in \cite{simpson2017} forms the basis of this venture. \\ \\
We derived the PC prior for the shape parameter of the Weibull model and investigated its performance with respect to the a priori belief about the applicability of the base model. Some popular priors were also investigated, and some surprising features were discovered, including their inability to contract to the base model for most hyperparameter values were illustrated. From these results we posit that the new PC prior is a reliable, transparent, principled and robust prior, even when prior information is weak or scarce.\\ \\
An application to data from a prostate cancer study invoking the proposed PC prior is presented (with a sensitivity study) to illustrate the use of this prior in various models (simple and more complicated), like the joint survival-longitudinal model. \\
This paper provides the user with a principled prior for the Weibull model that behaves in a predictable and reliable way, also in terms of its contraction to the exponential base model, even with little or no prior information. It was shown to be robust with regards to the hyperparameter specification and we advocate the use of the PC prior in forthcoming applications of the Weibull model.

\bibliographystyle{spmpsci}      
\bibliography{BioJ.bib}   

\end{document}